\documentclass{article}

\usepackage{arxiv}

\usepackage[utf8]{inputenc} 
\usepackage[T1]{fontenc}    
\usepackage{hyperref}       
\usepackage{url}            
\usepackage{booktabs}       
\usepackage{amsfonts}       
\usepackage{nicefrac}       
\usepackage{microtype}      
\usepackage{lipsum}
\usepackage{graphicx}
\graphicspath{ {./images/} }
\usepackage{amsmath}
\usepackage{amssymb}

\title{The X-ray photon index–Eddington ratio relation in radio-quiet quasars from XMM-Newton and SDSS}

\author{
 SH. M. Shehata \\
  LUX, CNRS UMR 8262 \\
  Observatoire de Paris \\ 
  61 Avenue de l'Observatoire, 75014 France\\
  National Research Institute of Astronomy and Geophysics \\
  Astronomy Department \\
  Cairo, 11421, Egypt \\
  \texttt{Sherehan.Shehata@obspm.fr} \\
   \And
 Baraa Hany \\
  Science Faculty, Helwan University \\
  Physics Department \\
  Cairo, 11795, Egypt \\
  \texttt{} \\
  \And
 Reham Mostafa \\
  Science Faculty, Fayoum University \\ 
  Physics Department \\
  Fayoum 63514, Egypt \\
  \texttt{rma04@fayoum.edu.eg} \\
}

\begin{document}
\maketitle
\begin{abstract}
This study presents a comprehensive X-ray spectroscopic analysis of 642 quasars, obtained by cross-matching the XMM-Newton Serendipitous Source Catalog (DR11) with the Sloan Digital Sky Survey (DR16) quasar catalog. After stringent quality filtering and automated spectral reduction, we derived reliable photon indices ($\Gamma$) and intrinsic 2--10 keV X-ray luminosities. Using multiwavelength data, sources were classified into 561 radio-quiet (RQ) and 81 radio-loud (RL) quasars. We estimate the bolometric luminosity and Eddington ratio ($\lambda_{\mathrm{Edd}}$) from absorption-corrected X-ray measurements and virial black hole masses. Our primary objective is to establish and characterize the fundamental relationship between the photon index and Eddington ratio  for RQ population. We find that RQ quasars exhibit systematically higher Eddington ratios, peaking at $\log \lambda_{\mathrm{Edd}} \approx -1.2$, and softer spectra with $\Gamma \approx 2.0$. A statistically significant positive correlation between $\Gamma$ and $\lambda_{\mathrm{Edd}}$ is detected in RQ quasars, supporting disk--corona coupling models. To validate our results within the broader context of AGN evolution, we further examine the dependence of $\lambda_{\mathrm{Edd}}$ on redshift ($z$) and black hole mass ($M_{\mathrm{BH}}$). For RQ quasars, $\lambda_{\mathrm{Edd}}$ increases with redshift and decreases with $M_{\mathrm{BH}}$, in strong agreement with recent results \cite{aggarwal2024evidence}, highlighting the universal nature of these accretion trends. By correlating spectral slope with accretion rate, this work provides new insights into the interplay between accretion physics, jet activity, and the cosmic evolution of quasars.
\end{abstract}


\section{Introduction}
Active galactic nuclei (AGNs) are powerful astrophysical objects that are driven by accretion onto supermassive black holes (SMBHs) located at the core of their host galaxy. They emit an enormous amount of non-stellar radiation, which has been detected in the radio, microwave, infrared (IR), optical, ultraviolet (UV), X-ray, and gamma-ray wave bands. Based on their radio power, AGN have been divided into two classes radio-quiet and radio-loud (e.g. \cite{Baum&Heckman, MillerRawlings}). Radio-loud AGN are capable of launching powerful relativistic jets that emit synchrotron radiation and dominate the radio band, and they constitute approximately $10\%$ of the overall AGN population (e.g.\cite{Begelman}). The radio emission is typically $10^3$ times brighter than in the radio-quiet AGN \cite{panessa}. Compared to the radio-quiet AGN, the radio-loud AGNs were found to exhibit harder X-ray spectra and higher X-ray luminosities on average \cite{Gupta}.

The X-ray spectrum of AGN exhibits an intrinsic power-law spectrum which is often described by a parameter known as the photon index ($\Gamma$). Simply, the photon index quantifies the slope of the X-ray spectrum: a steeper slope (larger $\Gamma$) indicates a 'softer' spectrum with more low-energy photons, while a flatter slope (smaller $\Gamma$) signifies a 'harder'spectrum with a greater proportion of high-energy photons. For AGN, the photon index typically ranges from approximately 1.4 to 2.8. The distribution of this spectrum in local AGN can be approximated by a Gaussian with mean 1.95 and standard deviation 0.15 \cite{Nandra,Bianchi}. This spectral structure can be explained by the inverse Compton scattering of accretion disk photons by a heated corona of relativistic electrons \cite{Titarchuk,Done}.This suggests a link between the X-ray corona and the accretion disks for AGNs \cite{haardt,Lusso_2016}.

Understanding the physical mechanisms governing these powerful cosmic engines requires careful analysis of their observable properties (see \cite{Laha}). Among the most crucial of these properties are the X-ray photon index and the Eddington ratio, which serve as key diagnostics for probing the accretion physics and the intrinsic nature of AGN. The Eddington ratio ($\lambda_{\mathrm{Edd}}$ or L/$L_{\mathrm{Edd}}$) is a fundamental parameter in accretion physics, representing the ratio of an AGN's bolometric luminosity ($L_{\mathrm{bol}}$, the total energy output across all wavelengths) to its Eddington luminosity ($L_{\mathrm{Edd}}$). Numerious studies have revealed the correlation between the X-ray photon index and the Eddington ratio in AGN. Early studies found a positive correlation between Eddington ratio and photon index (e.g.\cite{Lu, Wang, Porquet}). Several subsequent studies have reported results consistent with this relationship. Positive correlations between the X-ray photon index and the Eddington ratio have been found in samples of luminous quasars \cite{Shemmer2006,shemmer2008}, in a large quasar sample spanning $0.053 \leq z \leq 4.2$ \cite{risaliti2009}, and in a sample of 69 X-ray--selected quasars at $0.5 \leq z \leq 2.0$ with $42.5 \leq \log(L_{\mathrm{X}}) \leq 45.5$ \cite{brightman2013}. Similar results were also reported by \cite{kawamuro} for a sample of 45 local, moderately obscured (Compton-thin) AGNs observed with \textit{Suzaku} and \textit{Swift}/BAT. 

In contrast, \cite{Trakhtenbrot2017_BASS_Gamma_Ledd} reported a statistically significant but very weak correlation between $\Gamma$ and $L/L_{\mathrm{Edd}}$ in a sample of 228 hard X-ray--selected, low-redshift AGNs. Moreover, several recent studies have found no significant correlation between $\Gamma$ and $L/L_{\mathrm{Edd}}$ in samples of low-accreting AGNs \cite{Jana} as well as in samples of high-$\lambda_{\mathrm{Edd}}$ AGNs \cite{Laurenti2022,Laurenti}. The possible interpretation of this correlation according to some theories (e.g.\cite{Fabian2015,Yang2015,Kara2017,Ricci2018_AGN_Corona,Barua2020_Ark564}) is that at higher accretion rates, stronger UV/optical emission from the accretion disk enhances Compton cooling in the X-ray corona, lowering its temperature and/or optical depth and thereby producing a softer X-ray spectrum. On the contrary, some studies have reported an anti-correltion between $\Gamma$ and L/$L_{\mathrm{Edd}}$, particularly at very high or very low Eddington ratios, or in specific types of AGN (see \cite{GuCao2009_LLAGN,Younes2011_LINERs,JangGliozzi2014_LowAccretingAGN,Yang2015,kawamuro}). \cite{GuCao2009_LLAGN} suggested that the negative correlation can be explained by a RIAF model, in which a higher accretion rate leads to a hotter, denser plasma that more efficiently up-scatters seed photons to higher energies, thus producing a harder X-ray spectrum \cite{Narayan}.

Investigating Eddington ratio’s dependence on redshift is crucial for understanding the growth of supermassive black holes (SMBHs) and the evolution of galaxies. Previous studies have tried to investigate the evolution of the Eddington ratio with the redshift and black hole mass. \cite{NetzerTrakhtenbrot2007} analyzed thousands of SDSS  type1 AGN at z $\leq 0.75$ and confirmed that the Eddington ratio is smaller for larger mass black holes at all redshifts whereas found evidence that the peak of Eddington ratio distribution shifts to lower values at lower redshifts. Evidence for cosmic downsizing, where the number density of quasars peaks at higher redshift with increasing black hole mass, has been found in a sample of 9886 SDSS quasars at \(1 < z < 4.2 \) \cite{Kelly}. Recent study have revealed a complex relationship between $\lambda_{\mathrm{Edd}}$ and z in a large sample of 132000 AGNs at \( 0.1 < z < 2.4 \) \cite{aggarwal2024evidence}. He found that for similar-size SMBHs, $\lambda_{\mathrm{Edd}}$ decreases as z decreases, and that for a given redshift, larger SMBHs have a lower $\lambda_{\mathrm{Edd}}$. 

In this paper, our objective is to examine the relation between $\Gamma$ and $\lambda_{\mathrm{Edd}}$ in RQ quasar sample. In addition, we test how relevant the evolution of the Eddington ratio with the redshift is in different SMBH mass for RQ quasar sample. This work is organized as follows. In section 2, we present spectral analysis of our initial quasar sample and their classification into radio-loud (RL) and radio-quiet (RQ) objects. In section 3, we estimate the bolometric luminosity and the corresponding Eddington ratio for each source in RQ sample. In section 4, we investigate the correlation between the photon index $\Gamma$ and Eddington ratio $\lambda_{\mathrm{Edd}}$ for RQ quasars. We show the dependence of Eddington ratio $\lambda_{\mathrm{Edd}}$ on redshift (z) and black hole mass $M_{\mathrm{BH}}$ in section 5. Finally, we summarize and discuss our results in section 6.

\section{X-ray properties and spectral analysis}
\label{sec:headings}
 We constructed our X-ray sample by cross-matching point-like sources from the XMM-Newton Serendipitous Source Catalog Data Release 11 (4XMM--DR11, \cite{rosen2016}) with quasars from the Sloan Digital Sky Survey (SDSS). The SDSS Data Release 16 (DR16) quasar catalog \cite{lyke2020, wu2022catalog} was used as the parent catalog for the positional cross-matching, while physical quantities such as optical continuum flux densities and virial black hole masses were adopted from the SDSS Data Release 7 (DR7) quasar catalog \cite{shen2011} to ensure homogeneous and consistently derived spectral measurements.
 
 \subsection{Sample}
We first selected X-ray sources from the 4XMM--DR11 catalog classified as point-like, based on the EPIC extent parameter. Sources with EPIC extent $< 6$ were considered point-like. This initial selection yielded 811,211 X-ray sources. We then applied the selection criteria of \cite{shehata2021redshift}, with the exception that we selected a minimum of 800 total counts per source in the EPIC detectors to ensure sufficient spectral quality for reliable spectral fitting. This reduced the sample to 20,232 sources. To ensure high data quality, we keep only sources with \texttt{FLAG = 0} in the 4XMM catalog, corresponding to detections free from known data quality issues, high background contamination, or calibration problems. After applying this criterion, 8,904 X-ray sources remained. The final X-ray catalog contains 6,353 unique sample after excluding repeated detections and choosing the highest counts for each source. The optical quasar sample was taken from SDSS--DR16, which provides spectroscopically confirmed quasars with a reported completeness of 99.8\% and a contamination rate between 0.3\% and 1.3\%. The catalog contains 750,414 quasars, including 225,082 newly identified sources. We excluded objects classified as blazars, galaxies, or stars, yielding a final optical sample of 718,850 quasars. We cross-matched the filtered X-ray and optical catalogs using a positional search radius of 5 arcseconds obtaining 799 X-ray--optical source.
   
\subsection{X-ray reduction and spectral fitting}
  
We performed a systematic reduction and analysis of the XMM observations using the most recent version of the \texttt{Science Analysis System} (SAS) \cite{2017xru..conf...84G}. For each source, we obtained the Galactic hydrogen column density (\(N_{\mathrm{H,Gal}}\)) from the \texttt{NH} Tool \footnote{https://heasarc.gsfc.nasa.gov/cgi-bin/Tools/w3nh/w3nh.pl}, which interpolates over high-resolution H\textsc{i} surveys.
   
We excluded observations affected by background flaring, SAS processing or calibration warnings, sources located at the edge of the EPIC field of view or affected by source overlap, as well as EPIC-pn timing mode observations. After applying these criteria, the sample was reduced from 799 to 777 sources. Spectral fitting was then performed for each source using Xspec software package \cite{arnaud1996} within  0.3–10.0 keV energy range. The baseline model for all sources consisted of a power law modified by Galactic absorption (tbabs), with the (\(N_{\mathrm{H,Gal}}\)) value fixed to each source\'s line of sight. From these baseline fits, we derived the photon index (\(\Gamma\)) and the intrinsic (absorption-corrected X-ray) luminosity in the 2--10~keV band (\(L_{\mathrm{X}}\)).

For sources where simple spectral models (power law or power law plus blackbody) resulted in poor fits, indicated by reduced C-statistic values greater than 1.5, we tested the presence of additional intrinsic absorption by including a redshifted absorption component (ztbabs). In several cases, this additional component significantly improved the quality of the spectral fits. The entire reduction and analysis pipeline was fully automated using a combination of shell and Tcl scripting to ensure consistency across the sample.

   \begin{figure*}
\centering
\includegraphics[width=17cm]{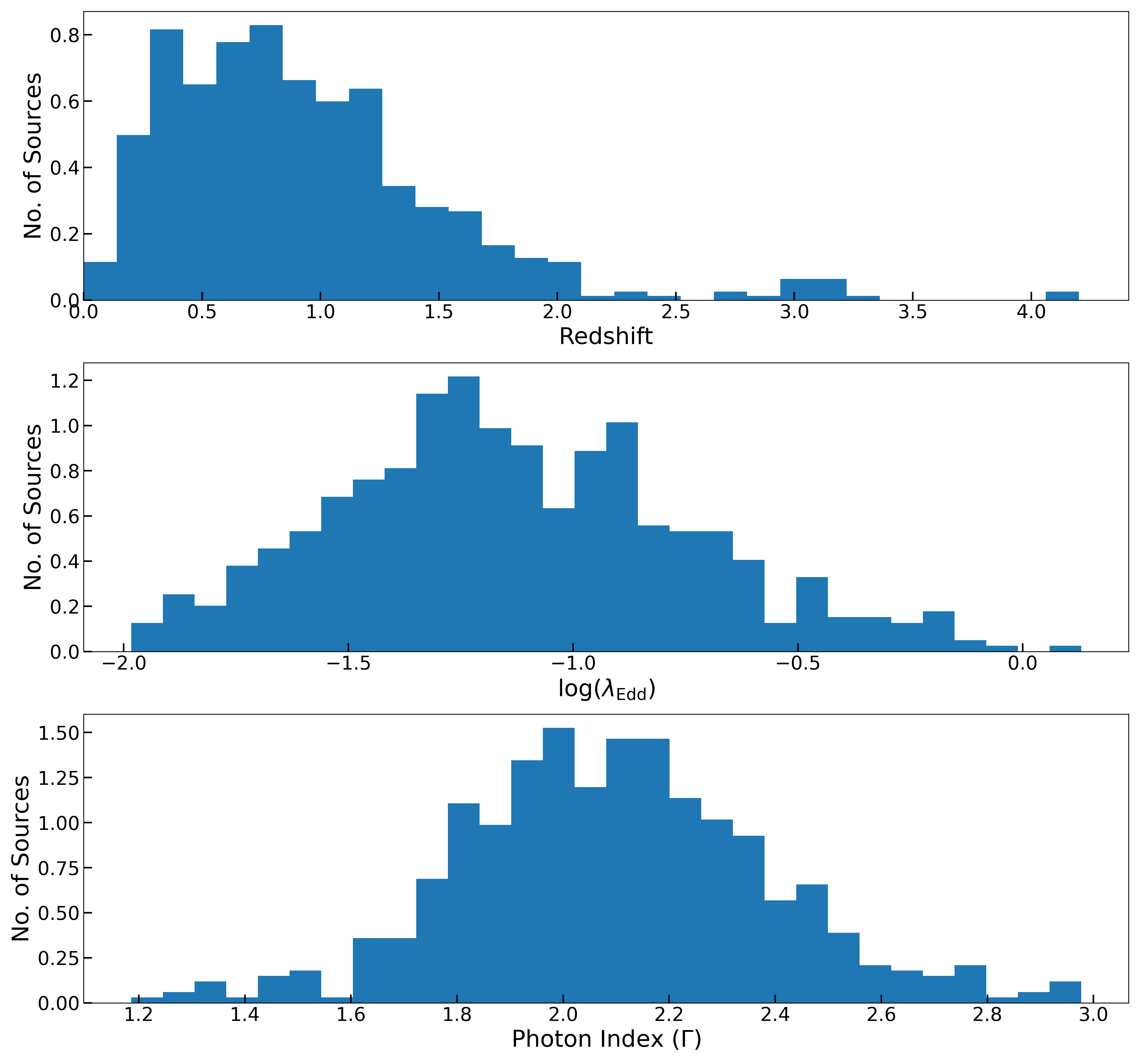}
    \caption{Distributions of redshift ($z$; top panel), Our sample comprises 541 radio-quiet active galactic nuclei (AGN) with redshifts spanning from z = 0.053 to z = 4.2 (median z = 0.82). The distribution across redshift bins reveals that most sources $(\sim63\%)$ lie below z = 1.0,logarithm of the Eddington ratio ($\log \lambda_{\rm Edd}$, middle panel), and X-ray photon index ($\Gamma$, bottom panel) for the radio-quiet quasar samples. The RQ population peaks at a higher accretion rate ($\log \lambda_{\rm Edd} \approx -1.2$) and exhibits a softer spectrum ($\Gamma \approx 2.0$)}
    \label{fig1}
\end{figure*}

\subsection{Radio classification of the quasar sample}
   We classified our quasar sample into radio-loud (RLQ) and radio-quiet (RQQ) objects using the radio-loudness parameter $R$, defined as the ratio of the rest-frame 5~GHz radio flux density to the rest-frame optical flux density at 2500~\AA, following \cite{jiang2007radio}:
\begin{equation}
    \label{eq1}
    R = \frac{f_{\mathrm{5\,GHz}}}{f_{2500\text{\AA}}},
\end{equation}
where $f_{\mathrm{5\,GHz}}$ is the rest-frame radio flux density at 5~GHz and $f_{2500\text{\AA}}$ is the rest-frame optical flux density at 2500~\AA  \cite{Kellermann1989}.

Radio flux densities at 1.4~GHz were obtained from the Faint Images of the Radio Sky at Twenty-Centimeters (FIRST) survey \cite{Becker1995} and converted to 5~GHz assuming a radio spectral index of $\alpha = -0.5$, such that $f_{\nu} \propto \nu^{\alpha}$.

Optical continuum flux densities at 2500~\AA\ were adopted from the SDSS Data Release~7 (DR7) quasar catalog \cite{shen2011}, which provides uniformly modeled continuum measurements based on spectral fitting.

After cross-matching our X-ray selected sample with the FIRST and SDSS DR7 quasar catalogs, and retaining only sources with both radio and optical measurements, we obtained 642 quasars (135 sources were excluded due to missing measurements). From the computed $R$ values, sources were classified as radio-loud if $R > 10$ and radio-quiet if $R \leq 10$
\cite{Kellermann1989}. This resulted in a subsample of 81 RL quasars and 561 RQ quasars.

The small sample size of radio-loud quasars compared to radio-quiet quasars can lead to increased uncertainties in correlation analyses. Moreover, the X-ray emission of radio-loud quasars can be contaminated by jet-related processes, which may influence their observed properties. The comparatively small size of the RL sample is therefore expected, as radio-loud quasars constitute only $\sim$10\% of the overall quasar population. Thus, we restrict the remainder of our analysis to the radio-quiet quasar sample in order to obtain a clearer view of accretion-driven correlations.  

\section{Estimation of bolometric luminosity and Eddington ratio}

   To investigate the accretion properties of the active galactic nuclei (AGN) in our RQ sample, we determined the bolometric luminosity ($L_{\mathrm{bol}}$), Eddington luminosity ($L_{\mathrm{Edd}}$), and the corresponding Eddington ratio ($\lambda_{\mathrm{Edd}}$) for each source. The methodology is outlined below.

First, we derived the intrinsic absorption-corrected hard X-ray luminosity in the 2--10~keV band, $L_{\mathrm{X}}(2\text{--}10\,\mathrm{keV})$, from spectral analysis of the available XMM-Newton observations. This energy range is considered a reliable proxy of the coronal emission properties / accretion state, as it is less affected by contamination from the host galaxy or star formation processes \cite{brandt2005}.

Next, the bolometric luminosity was estimated by applying a bolometric correction ($k_{\mathrm{bol}}$) to the hard X-ray luminosity. This correction is essential for extrapolating from a specific energy band to the total radiative output across the entire electromagnetic spectrum \cite{elvis1994}. Following established prescriptions, we adopted the relation:
\begin{equation}
    \label{eq2}
L_{\mathrm{bol}} = k_{\mathrm{bol}} \times L_{\mathrm{X}}(2\text{--}10\,\mathrm{keV}).
\end{equation}

For this study, we used a constant X-ray bolometric correction factor of $k_{\mathrm{bol}} = 20$. This value is consistent with the luminosity-dependent prescriptions of \cite{marconi2004} for AGN with $L_{2\text{--}10\,\mathrm{keV}} \sim 10^{43\text{--}44}\,\mathrm{erg\,s^{-1}}$, and with the typical X-ray bolometric corrections reported by \cite{vasudevan2007} for radiatively efficient AGN. The choice is further supported by recent large-sample studies, which report a mean X-ray bolometric correction of $\sim 20$, albeit with a dispersion of approximately one order of magnitude \cite{Duras2020}.

The black hole masses ($M_{\mathrm{BH}}$) for our sample were obtained from the comprehensive Sloan Digital Sky Survey (SDSS) catalog compiled by \cite{shen2011}. In this catalog, the masses of virial black holes are estimated from the widths of broad emission lines and the continuum luminosity, a standard technique for large AGN samples \cite{vestergaard2006}.

Using these black hole masses, we calculated the Eddington luminosity for each source. The Eddington luminosity represents the theoretical maximum luminosity that a body can achieve when there is a balance between the outward force of radiation and the inward gravitational force \cite{eddington1926}. It is defined by the standard formula:
\begin{equation}
    \label{eq3}
L_{\mathrm{Edd}} = 1.26 \times 10^{38} \left( \frac{M_{\mathrm{BH}}}{M_\odot} \right) \, \mathrm{erg\,s^{-1}}.
\end{equation}

Finally, we computed the Eddington ratio ($\lambda_{\mathrm{Edd}}$), which is a crucial parameter for understanding the accretion state of the supermassive black hole. It is defined as the ratio of the bolometric luminosity to the Eddington luminosity:
\begin{equation}
    \label{eq4}
\lambda_{\mathrm{Edd}} = \frac{L_{\mathrm{bol}}}{L_{\mathrm{Edd}}}.
\end{equation}

\begin{figure}[ht!]
    \centering
        \includegraphics[width=\hsize]{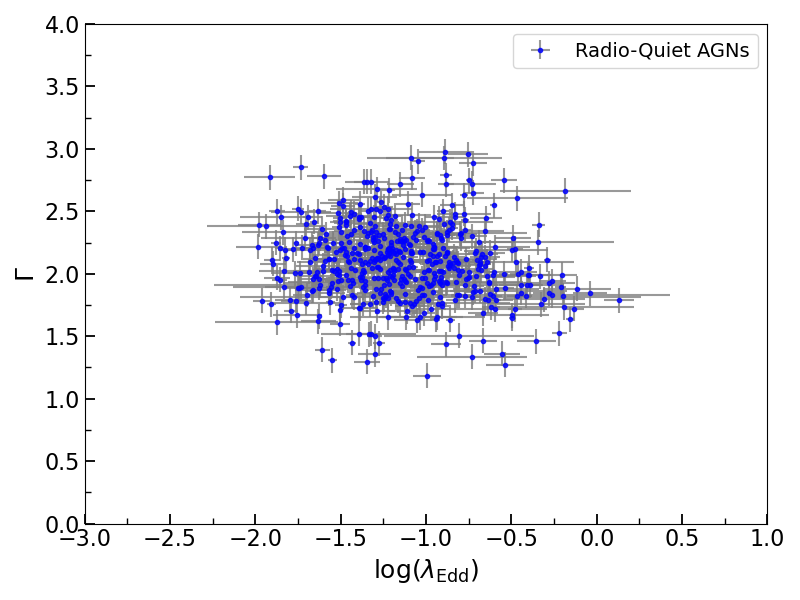}
    \caption{The relationship between the X-ray photon index ($\Gamma$) and the Eddington ratio ($\lambda_{\mathrm{Edd}}$) for the radio-quiet sample.}
    \label{fig2}
\end{figure}

This ratio provides a direct measure of accretion efficiency relative to the Eddington limit, which is fundamental for characterizing the physical processes that govern the activity of AGN \cite{kollmeier2006, fabian2012}.

\section{X-ray spectral index and Eddington ratio}

Before testing for correlations between the spectral and accretion properties of our sources, we first characterize the overall distributions of the radio-quiet sample. Figure~\ref{fig1} presents distributions of redshift ($z$), the logarithm of the Eddington ratio ($\log \lambda_{\mathrm{Edd}}$), and the X-ray photon index ($\Gamma$) for RQ population. The top panel of Figure~\ref{fig1} shows that the distribution peaks at $z \approx 0.8$. The distribution of the Eddington ratio (middle panel) is systematically shifted towards higher accretion rates, with their distribution peaking at $\log \lambda_{\mathrm{Edd}} \approx -1.2$. As shown in the bottom panel the radio-quiet sources exhibit softer spectra, i.e., larger photon indices compared to radio-loud quasars, with their photon index distribution peaking at $\Gamma \approx 2.0$. Given these fundamental properties of the RQ population, we proceed in the next section to formally check the relationship between the spectral index and Eddington ratio. We examined the correlation between the X-ray photon index ($\Gamma$) and the Eddington ratio ($\lambda_{\mathrm{Edd}}$) for the radio-quiet sample (see Figure~\ref{fig2}). We excluded two sources from our analysis because the large uncertainties in their spectral measurements prevented any significant constraints on their luminosities, and then performed a Spearman rank-order correlation test. The result yielded a correlation coefficient of $\rho = 0.12$ with a p-value of $4.3 \times 10^{-3}$, indicating a weak but statistically significant positive monotonic correlation.

This trend is physically consistent with expectations from accretion disk-corona models, where higher accretion rates lead to enhanced Compton cooling of the corona, resulting in softer X-ray spectra (i.e., larger photon indices). Similar positive correlations between $\Gamma$ and $\lambda_{\mathrm{Edd}}$ have been reported in previous studies of radio-quiet quasars. For example, \cite{shemmer2008} found $\rho \sim 0.3$ for a sample of high redshift SDSS quasars, while \cite{risaliti2009} and \cite{brightman2013} reported comparable results using XMM-Newton selected samples. Although our correlation strength is somewhat weaker, the larger size of our sample lends statistical robustness to the observed trend.

\begin{figure}[ht!]
    \centering
        \includegraphics[width=\hsize]{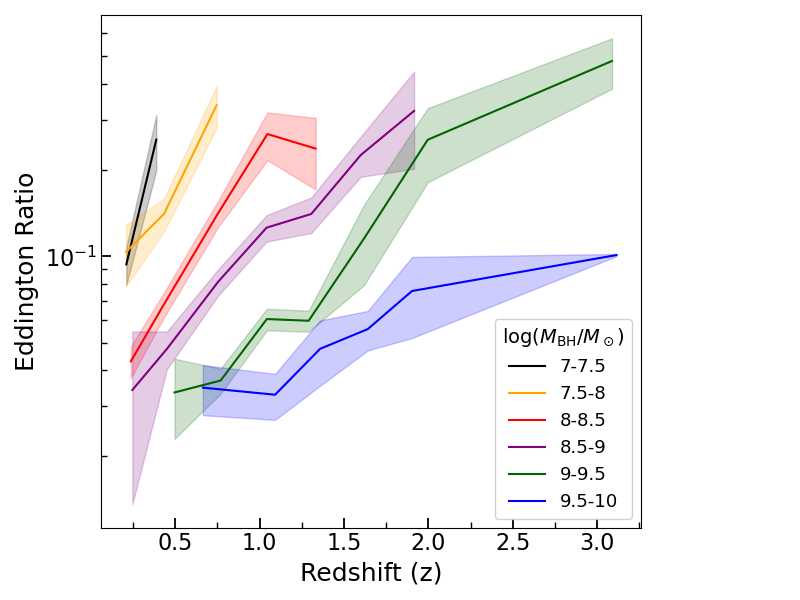}
    \caption{Eddington Ratio vs. Redshift by Black Hole Mass for radio-quiet AGN. This plot displays the Eddington ratio ($\lambda_{Edd}$) as a function of redshift (z) for various supermassive black hole (SMBH) mass bins, specifically for radio-quiet AGN. The plot indicates an increase in $\lambda_{Edd}$ with increasing z and a decrease in $\lambda_{Edd}$ with increasing $M_{BH}$ at a fixed redshift.}
    \label{fig3}
\end{figure}

To assess the robustness of this result, we repeated the analysis by propagating the uncertainties. In this case, the correlation coefficient decreases to $\rho = 0.106$, with a corresponding p-value of $1.24 \times 10^{-2}$. This reduction in both the correlation strength and statistical significance indicates that the observed trend is sensitive to measurement uncertainties and intrinsic scatter. Consequently, while the best-fit trend remains positive, the data are also consistent with a very weak or absent correlation.

\section{Eddington ratio evolution with redshift}

   We investigated the relationship between the Eddington ratio and the redshift for RQ sample (Figure~\ref{fig3}). Sources span the redshift range \( 0.053 < z < 4.2 \) and are more densely populated at \( z < 2 \). We observe a slight increase in the mean \(\log(\lambda_{\mathrm{Edd}})\) with redshift, although the scatter remains large at all redshifts. This trend is consistent with the idea that quasars at higher redshifts tend to accrete at higher fractions of their Eddington limit, possibly reflecting an era of more efficient black hole growth in the early Universe \cite{Trakhtenbrot2012, shen2011}. Our analysis of the Eddington ratio ($\lambda_{Edd}$) as a function of redshift (z) and supermassive black hole (SMBH) mass ($M_{BH}$) reveals a clear dependence on both parameters. This is in consistent with previous results in the literature. \cite{SchulzeWisotzki2010} showed that at low redshift (z < 0.3) the Eddington ratio distribution peaks around $\lambda_{Edd}$ \(\sim\) 0.1, and there was a strong decline in the fraction of highly accreting AGNs with increasing black hole mass. It was found that the Eddington ratio increases with redshift for type 1 and type 2 AGN at any given $M_{BH}$ \cite{Lusso}. Moreover, \cite{Trakhtenbrot2012} observed a steep rise of Eddington ratio with redshift up to z $\sim 1$ followed by a flattening at $\lambda_{Edd}$ $\sim 1$ and sources with lower $M_{BH}$ consistently show higher $\lambda_{Edd}$ at all redshifts. Recent and direct investigation using a huge sample of 132,000 SMBHs have shown that for similar-size SMBHs, $\lambda_{\mathrm{Edd}}$ decreases as z decreases, and that for a given redshift, larger SMBHs have a lower $\lambda_{\mathrm{Edd}}$ \cite{aggarwal2024evidence}.

As depicted in Figure~\ref{fig3}, We adopted equal-width redshift bins spanning the full redshift range of the sample ($0.053 \le z \le 4.201$), using 14 bins of width $\Delta z \simeq 0.30$. This choice provides uniform coverage of cosmic time while allowing a consistent comparison across different black hole mass bins. The number of sources per bin varies substantially due to the flux-limited nature of the sample, ranging from $\sim$75--140 objects at $z \lesssim 1$ to fewer than 10 objects per bin at $z \gtrsim 2$. Bins with very low statistical significance are not shown in the figure. We observe that for a given SMBH mass, the Eddington ratio generally increases with increasing redshift. This trend is in excellent agreement with \cite{aggarwal2024evidence}, who reported that the Eddington ratio decreases as redshift decreases for SMBHs of similar mass. This suggests a cosmic evolution in the accretion properties of black holes, with higher accretion rates (relative to the Eddington limit) being more prevalent at earlier cosmic epochs.

Furthermore, our results for radio-quiet AGN demonstrate a strong inverse correlation between the Eddington ratio and the black hole mass at any given redshift. As illustrated in Figure~\ref{fig3}, higher mass SMBHs consistently exhibit lower Eddington ratios compared to their lower mass counterparts at the same redshift. This observation directly supports the conclusion of \cite{aggarwal2024evidence} that larger SMBHs tend to have lower Eddington ratios. This implies that the efficiency of accretion might be intrinsically linked to the black hole\'s mass, with more massive black holes accreting at a smaller fraction of their Eddington limit.

\section{Summary and discussion}
In this study, we provide a robust, large-scale statistical characterization of the relationship between X-ray emission and accretion physics in Active Galactic Nuclei (AGN). We validate the standard model where the accretion disk and the X-ray corona are coupled, specifically testing if higher accretion rates lead to more efficient cooling of the corona. We construct an X-ray sample of 642 quasars out of cross-matching the X-ray catalog XMM-DR11 with the quasar catalog SDSS-DR16. The X-ray spectra of our sample are fitted with an absorbed power-law model to estimate the photon index ($\Gamma$) and the intrinsic X-ray luminosity in the 2 – 10 keV band. We classify our quasar sample into 81 radio-loud (RL) and 561 radio-quiet (RQ) objects. For the radio-quiet quasar sample, we determined the bolometric luminosity ($L_{\mathrm{bol}}$), Eddington luminosity ($L_{\mathrm{Edd}}$), and the corresponding Eddington ratio ($\lambda_{\mathrm{Edd}}$) for each source. The Eddington ratio $\lambda_{Edd}$ in our radio-quiet quasar sample ranges from $10^{-2}$ to 0.9 peaking at $6 \times 10^{-2}$ . 

Our analysis of 561 quasars reveals a statistically significant, albeit weak, positive correlation between $\Gamma$ and $\lambda_{Edd}$, with a Spearman rank coefficient of r = 0.12 and p = $4 \times 10^{-3}$. The statistical significance, driven by our large sample size, suggests an underlying physical link, consistent with the standard disk-corona model where more efficient cooling from a stronger accretion disk leads to a softer X-ray spectrum. A significant correlation between $\Gamma$ and $\lambda_{Edd}$ in 25 moderate-luminosity radio-quiet AGNs at z < 0.5 was found by \cite{Shemmer2006} with a strong correlation coefficient of 0.60 and probability of $1.6 \times 10^{-3}$. Moreover, \cite{shemmer2008} introduced 10 highly luminous AGNs at z = 1.3 -- 3.2 to the 25 moderate-luminosity radio-quiet AGNs sample and found that $\Gamma$--$\lambda_{Edd}$ correlation remained significant, but the correlation coefficient decreased to 0.55 and the probability dropping from p = $1.6 \times 10^{-3}$ to p = $4 \times 10^{-4}$. In addition, \cite{risaliti2009} presented highly statistically significant correlation with a linear correlation coefficient of 0.32, for a sample of 343 AGNs from SDSS–XMM-Newton quasar survey. They show that this correlation become stronger for the subsample of objects with black hole masses determined from the H$\beta$ line, and weaker (but still significant) for those based on Mg $\mathrm{II}$. On the other hand, \cite{Trakhtenbrot2017_BASS_Gamma_Ledd} showed a weak (but statistically significant) correlation across a primary sample of 228 AGN, despite a large amount of scatter. Furthermore, When examining subsamples of AGN that differ in the method used to compute $M_{\mathrm{BH}}$, they found either no or weak correlation. This correlation appears to break down at the highest accretion rates. Recent studies have reported lack of correlation between $\Gamma$ and $\lambda_{Edd}$ for high $\lambda_{Edd}$ quasars ($L_{bol}$ $\sim$ $\times$ $10^{46}$ erg $s^{-1}$) \cite{Laurenti2022,Laurenti}. This suggests the emergence of yet another accretion regime, where immense radiation pressure from the disk may fundamentally alter the structure and efficiency of the disk-corona system.

Our results describes the "softer-when-brighter" mechanism. As accretion increases, the disk produces more ultraviolet/soft photons that pass through the corona and "cool" it via inverse Compton scattering, which results in a higher (softer) photon index $\Gamma$ \cite{shemmer2008,sobolewska2009} and can be explained by Comptonization models where the plasma temperature is regulated by the supply of soft photons from the disk \cite{Ricci2018_AGN_Corona}.

We examined the dependence of $\lambda_{\mathrm{Edd}}$ on black hole mass ($M_{\mathrm{BH}}$) and redshift for our radio-quiet quasar sample. We find two distinct trends: (1) for black holes of comparable mass, $\lambda_{\mathrm{Edd}}$ decreases toward lower redshifts, and (2) at a given redshift, sources with larger black hole masses tend to exhibit lower $\lambda_{\mathrm{Edd}}$. These findings are broadly consistent with previous observational studies in the literature (e.g.\cite{NetzerTrakhtenbrot2007,SchulzeWisotzki2010,Lusso}), and has been confirmed by recent work using independent, large sample \cite{aggarwal2024evidence}. However, we emphasize that the interpretation of these observed trends requires significant caution, as they are influenced by several inherent biases. First, as the Eddington ratio is defined by the black hole mass ($\lambda_{\mathrm{Edd}} \propto L/M_{\mathrm{BH}}$), any analysis of the $\lambda_{\mathrm{Edd}}$--$M_{\mathrm{BH}}$ plane is affected by mathematical coupling. More importantly, as demonstrated through detailed forward-modeling, a flux limit will naturally produce an apparent anti-correlation between black hole mass and Eddington ratio, because at any given luminosity, only low-mass, high-$\lambda_{\mathrm{Edd}}$ objects or high-mass, low-$\lambda_{\mathrm{Edd}}$ objects are selected \cite{SchulzeWisotzki2010}. This effect can dominate any intrinsic physical correlation. Furthermore, \cite{Steinhardt} identified boundaries of the observed quasar distribution in the Mass-Luminosity plane, including the Sub-Eddington boundary, which complicates the interpretation of simple statistical trends. The complex interplay of these biases with scatter in virial mass estimates can create the illusion of strong cosmic evolution, such as an increase in the average $\lambda_{\mathrm{Edd}}$ with redshift, simply because lower-luminosity populations become progressively undetectable at earlier cosmic times (for a review, see \cite{Shen}). A full forward-modeling analysis to deconvolve these selection effects is beyond the scope of this work. Therefore, while our results are qualitatively consistent with the 'downsizing' evolutionary scenario in which the most massive black holes experienced their peak growth at earlier cosmic epochs we do not claim them as independent, definitive evidence for the intrinsic nature of black hole growth. Our results should be interpreted as an independent data point that, within these limitations, aligns with the current understanding of AGN evolution.

\section*{Acknowledgements}
      The authors thank Johannes Buchner for helpful discussions and for providing helpful comments. SH. M. Shehata acknowledges the Science, Technology \& Innovation Funding Authority (STDF) and the Institut Français d\'Egypte (IFE) for support through the STDF–IFE postdoc fellowship program (Call 11), in partnership with the Embassy of France in Egypt. SH. M. Shehata also expresses sincere gratitude to the LUX Laboratory, CNRS UMR 8262, Observatoire de Paris, for hosting this research during the postdoctoral fellowship period. This work is based on observations obtained with XMM-NEWTON, an ESA science mission with instruments and contributions directly funded by ESA member states and NASA. Funding for SDSS-III has been provided by the Alfred P. Sloan Foundation, the Participating Institutions, the National Science Foundation, and the  U.S. DOE Office of Science.

\bibliographystyle{unsrt}  
\bibliography{references} 

@article{marconi2004,
  author = {Marconi, A. and Risaliti, G. and Gilli, R. and Hunt, L. K. and Maiolino, R. and Salvati, M.},
  title = {Local supermassive black holes, relics of active galactic nuclei and the {X-ray} background},
  journal = {Monthly Notices of the Royal Astronomical Society},
  volume = {351},
  number = {1},
  pages = {169--185},
  year = {2004},
  doi = {10.1111/j.1365-2966.2004.07765.x}
}

@article{vasudevan2007,
  author = {Vasudevan, R. V. and Fabian, A. C.},
  title = {Piecing together the {X-ray} background: bolometric corrections for active galactic nuclei},
  journal = {Monthly Notices of the Royal Astronomical Society},
  volume = {381},
  number = {4},
  pages = {1235--1251},
  year = {2007},
  doi = {10.1111/j.1365-2966.2007.12231.x}
}

@article{shen2011,
  author = {Shen, Yue and Richards, Gordon T. and Strauss, Michael A. and Hall, Patrick B. and Schneider, Donald P. and et al.},
  title = {A Catalog of Quasar Properties from {Sloan Digital Sky Survey Data Release} 7},
  journal = {The Astrophysical Journal Supplement Series},
  volume = {194},
  number = {2},
  pages = {45},
  year = {2011},
  doi = {10.1088/0067-0049/194/2/45}
}

@article{brandt2005,
  author = {Brandt, W. N. and Hasinger, G.},
  title = {Deep Extragalactic {X-ray} Surveys},
  journal = {Annual Review of Astronomy and Astrophysics},
  volume = {43},
  number = {1},
  pages = {827--859},
  year = {2005},
  doi = {10.1146/annurev.astro.43.051804.102213}
}

@article{elvis1994,
  author = {Elvis, Martin and Wilkes, Belinda J. and McDowell, Jonathan C. and Green, Richard F. and Bechtold, Jill and et al.},
  title = {Atlas of Quasar Energy Distributions},
  journal = {The Astrophysical Journal Supplement Series},
  volume = {95},
  pages = {1},
  year = {1994},
  doi = {10.1086/192093}
}

@article{vestergaard2006,
  author = {Vestergaard, M. and Peterson, B. M.},
  title = {Determining Central Black Hole Masses in Distant Active Galaxies and Quasars. {II. Improved} Virial Mass Estimates},
  journal = {The Astrophysical Journal},
  volume = {641},
  number = {2},
  pages = {689--709},
  year = {2006},
  doi = {10.1086/500572}
}

@book{eddington1926,
  author = {Eddington, A. S.},
  title = {The Internal Constitution of the Stars},
  publisher = {Cambridge University Press},
  year = {1926},
  address = {Cambridge},
  doi = {10.1017/CBO9780511694171}
}

@article{kollmeier2006,
  author = {Kollmeier, Juna A. and Onken, Christopher A. and Kochanek, Christopher S. and Gould, Andrew and Weinberg, David H. and et al.},
  title = {Black Hole Masses and {Eddington} Ratios at 0.3 < z < 4},
  journal = {The Astrophysical Journal},
  volume = {648},
  number = {1},
  pages = {128--139},
  year = {2006},
  doi = {10.1086/505646}
}

@article{fabian2012,
  author = {Fabian, A. C.},
  title = {Observational Evidence of Active Galactic Nuclei Feedback},
  journal = {Annual Review of Astronomy and Astrophysics},
  volume = {50},
  pages = {455--489},
  year = {2012},
  doi = {10.1146/annurev-astro-081811-125521}
}

@article{rosen2016,
  author = {Rosen, S. R. and Webb, N. A. and Watson, M. G. and Ballet, J. and Barret, D. and Braito, V. and Carrera, F. J. and Ceballos, M. T. and Coriat, M. and Della Ceca, R. and Denkinson, G. and Esquej, P. and Farrell, S. and Freyberg, M. J. and Grisé, F. and Guainazzi, M. and Heil, L. M. and Koliopanos, F. and Law-Green, J. and Lamer, G. and Lin, D. and Martino, R. and Michel, L. and Motch, C. and Nebot Gomez-Moran, A. and Page, C. G. and Page, M. J. and Page, K. L. and Pakull, M. W. and Pye, J. P. and Read, A. M. and Rodriguez, P. and Sakano, M. and Saxton, R. D. and Schwope, A. and Scott, A. E. and Sturm, R. and Traulsen, I. and Yershov, V. and Zolotukhin, I.},
  title = {{The XMM-Newton serendipitous survey. VII. The third XMM-Newton serendipitous source catalogue}},
  journal = {Astronomy \& Astrophysics},
  volume = {590},
  pages = {A1},
  year = {2016},
  doi = {10.1051/0004-6361/201526416}
}

@article{lyke2020,
  author = {Lyke, B. W. and Higley, A. N. and McLane, J. N. and Schurhammer, D. P. and Myers, A. D. and Ross, N. P. and Dawes, C. and Farr, W. and MacLeod, C. L. and Ivezic, Z. and Richards, G. T. and Strauss, M. A. and Dawson, K. S. and Weinberg, D. H. and AlSayyad, Y. and Brandt, W. N. and Denney, K. D. and Fan, X. and Hall, P. B. and Hennawi, J. F. and Paris, I. and Schneider, D. P. and White, M.},
  title = {{The Sloan Digital Sky Survey Quasar Catalog}: {Sixteenth Data Release} ({DR16Q})},
  journal = {The Astrophysical Journal Supplement Series},
  volume = {250},
  number = {1},
  pages = {8},
  year = {2020},
  doi = {10.3847/1538-4365/aba623}}

@article{arnaud1996,
  author = {Arnaud, K. A.},
  title = {XSPEC: The First Ten Years},
  journal = {Astronomical Data Analysis Software and Systems V},
  volume = {101},
  pages = {17},
  year = {1996}
}

@article{shemmer2008,
  author = {Shemmer, Ohad and Brandt, W. N. and Netzer, H. and Maiolino, R. and Kaspi, S.},
  title = {The Hard {X-ay} Spectrum as a Probe for Black Hole Growth in Radio-Quiet Active Galactic Nuclei},
  journal = {The Astrophysical Journal},
  volume = {682},
  number = {1},
  pages = {81--93},
  year = {2008},
  doi = {10.1086/588776}
}

@ARTICLE{risaliti2009,
       author = {{Risaliti}, G. and {Young}, M. and {Elvis}, M.},
        title = "{The Sloan Digital Sky Survey/XMM-Newton Quasar Survey: Correlation Between X-Ray Spectral Slope and Eddington Ratio}",
      journal = {The Astrophysical Journal Letters},
         year = 2009,
       volume = {700},
       number = {1},
        pages = {L6-L10},
          doi = {10.1088/0004-637X/700/1/L6},
}

@article{brightman2013,
  author = {Brightman, M. and Silverman, J. D. and Mainieri, V. and Ueda, Y. and Schramm, M. and Matsuoka, K. and Nagao, T. and Steinhardt, C. and Kartaltepe, J.},
  title = {A systematic study of the {X-ray} spectral properties of {AGN} in the {COSMOS} field},
  journal = {Monthly Notices of the Royal Astronomical Society},
  volume = {433},
  number = {2},
  pages = {2485--2500},
  year = {2013},
  doi = {10.1093/mnras/stt913}
}

@article{sobolewska2009,
    author = {Sobolewska, M. A. and Papadakis, I. E.},
    title = {The long-term X-ray spectral variability of AGN},
    journal = {Monthly Notices of the Royal Astronomical Society},
    volume = {399},
    number = {3},
    pages = {1597-1610},
    year = {2009},
    doi = {10.1111/j.1365-2966.2009.15382.x},
}

@article{Trakhtenbrot2012,
  author = {Trakhtenbrot, B. and Netzer, H.},
  title = {Black hole growth to z = 2 – {I. Improved} virial methods for measuring M$_{BH}$ and L/L$_{Edd}$},
  journal = {Monthly Notices of the Royal Astronomical Society},
  volume = {427},
  number = {4},
  pages = {3081--3102},
  year = {2012},
  doi = {10.1111/j.1365-2966.2012.22056.x}
}

@article{Kellermann1989,
  author = {Kellermann, K. I. and Sramek, R. and Schmidt, M. and Shaffer, D. B. and Green, R.},
  title = {{VLA observations of objects in the Palomar Bright Quasar Survey}},
  journal = {The Astronomical Journal},
  volume = {98},
  pages = {1195--1207},
  year = {1989},
  doi = {10.1086/115207}
}

@article{Becker1995,
  author = {Becker, R. H. and White, R. L. and Helfand, D. J.},
  title = {The {FIRST} {Survey}: Faint Images of the Radio Sky at Twenty-Centimeters},
  journal = {The Astrophysical Journal},
  volume = {450},
  pages = {559},
  year = {1995},
  doi = {10.1086/176166}
}

@article{aggarwal2024evidence,
  title={Evidence that {Eddington} ratio depends upon a supermassive black hole’s mass and redshift: implications for radiative efficiency},
  author={Aggarwal, Yash},
  journal={Monthly Notices of the Royal Astronomical Society},
  volume={530},
  number={2},
  pages={1512--1515},
  year={2024},
  doi={10.1093/mnras/stae914}
}

@article{Baum&Heckman,
  author = {Baum, Stefi A. and Heckman, Timothy M.},
  title = {Extended optical line emitting gas in powerful radio galaxies: Statistical properties and physical conditions},
  journal = {The Astrophysical Journal},
  volume = {336},
  pages = {681–708},
  year = {1989},
  doi = {10.1086/167048}
}

@article{MillerRawlings,
  author    = {P. Miller and S. Rawlings and R. Saunders},
  title     = {The radio and optical properties of the z < 0.5 {BQS} quasars},
  journal   = {Monthly Notices of the Royal Astronomical Society},
  year      = {1993},
  volume    = {263},
  pages     = {425--460},
  doi       = {10.1093/mnras/263.2.425}
}

@article{Begelman,
  author = {{Begelman}, Mitchell C. and {Blandford}, Roger D. and {Rees}, Martin J.},
  title   = {Theory of extragalactic radio sources},
  journal = {Reviews of Modern Physics},
  year    = {1984},
  volume  = {56},
  number  = {2},
  pages   = {255--351},
  doi     = {10.1103/RevModPhys.56.255}
}

@article{Gupta,
  author = {Gupta, Maitrayee, Marek Sikora and Katarzyna Rusinek.},
  title = {Comparison of {SEDs} of Very Massive Radio‑Loud and Radio‑Quiet {AGN}},
  journal = {Monthly Notices of the Royal Astronomical Society},
  volume = {492},
  pages = {315-325},
  year = {2020},
  doi = {10.1093/mnras/stz3384}
}

@article{Nandra,
  author = {Nandra, K. and K. A. Pounds},
  title = {Ginga Observations of the {X‑ray} Spectra of {Seyfert} Galaxies},
  journal = {Monthly Notices of the Royal Astronomical Society},
  volume = {268},
  pages = {405-429},
  year = {1994},
  doi = {10.1093/mnras/268.2.405}
}

@article{Bianchi,
  author = {{Bianchi, S.} and {Guainazzi, M.} and {Matt, G.} and {Fonseca Bonilla, N.} and {Ponti, G.}},
  title = {{CAIXA: A catalogue of AGN in the XMM‑Newton archive. I. Spectral analysis}},
  journal = {Astronomy \& Astrophysics},
  volume = {495(2)},
  pages = {421‑430},
  year = {2009},
  doi = {10.1051/0004-6361:200810620}
}

@article{Titarchuk,
  author = {Titarchuk, Lev},
  title = {Generalized Comptonization Models and Application to the Recent High‑Energy Observations},
  journal = {The Astrophysical Journal},
  volume = {434},
  pages = {570‑586},
  year = {1994},
  doi = {10.1086/174760}
}

@article{Lu,
  author = {Lu, Youjun and Qingjuan Yu},
title = {Two Different Accretion Classes in {Seyfert} 1 Galaxies and {QSOs}},
  journal = {The Astrophysical Journal},
  volume = {526},
  pages = {L5-L8},
  year = {1999},
  doi = {10.1086/312358}
}

@article{Wang,
doi = {10.1086/421906},
year = {2004},
volume = {607},
number = {2},
pages = {L107},
author = {Wang, Jian-Min and Watarai, Ken-Ya and Mineshige, Shin},
title = {The Hot Disk Corona and Magnetic Turbulence in Radio-quiet Active Galactic Nuclei: Observational Constraints},
journal = {The Astrophysical Journal},
}

@article{Porquet,
  author = {{Porquet, D.} and {Reeves, J. N.} and {O’Brien, P.} and {Brinkmann, W.}},
  title = {{XMM‑Newton} {EPIC} Observations of 21 Low‑Redshift {PG} Quasars},
  journal = {Astronomy \& Astrophysics},
  volume = {422(1)},
  pages = {85‑95},
  year = {2004},
  doi = {10.1051/0004‑6361:20047108}
}

@article{Shemmer2006,
  author = {Shemmer, Ohad and Brandt, W. N. and Netzer, Hagai and Maiolino, Roberto and Kaspi, Shai},
  title = {The Hard {X‑ray} Spectral Slope as an Accretion‑Rate Indicator in Radio‑Quiet Active Galactic Nuclei},
  journal = {The Astrophysical Journal},
  volume = {646},
  pages = {L29‑L32},
  year = {2006},
  doi = {10.1086/506911}
}

@article{Fabian2015,
  author = {Fabian, Andy C. and others},
  title = {Properties of {AGN} Coronae in the {NuSTAR} {Era}},
  journal = {Monthly Notices of the Royal Astronomical Society},
  volume = {451(4)},
  pages = {4375‑4383},
  year = {2015},
  doi = {10.1093/mnras/stv1218}
}

@article{Yang2015,
  author = {Yang, Q. X. and others},
  title = {Evidence for a Correlation between the {X-ray} Spectral Slope and {Eddington} Ratio in Low-Luminosity Active Galactic Nuclei},
  journal = {Monthly Notices of the Royal Astronomical Society},
  volume = {447(2)},
  pages = {1692–1702},
  year = {2015},
  doi = {10.1093/mnras/stu2535}
}

@article{Kara2017,
  author = {Kara, E. and others},
  title = {A Global Look at {X-ray} Time Lags in {Seyfert} Galaxies},
  journal = {Monthly Notices of the Royal Astronomical Society},
  volume = {468(3)},
  pages = {3489–3507},
  year = {2017},
  doi = {10.1093/mnras/stx670}
}

@article{Ricci2018_AGN_Corona,
  author       = {Ricci, C. and others},
  title        = {{BAT AGN Spectroscopic Survey ‒ XII. The relation between coronal properties of active galactic nuclei and the Eddington ratio}},
  journal      = {Monthly Notices of the Royal Astronomical Society},
  year         = {2018},
  volume       = {480(2)},
  pages        = {1819-1830},
  doi          = {10.1093/mnras/sty1879},
}

@article{Barua2020_Ark564,
  author       = {Barua, Samuzal and others},
  title        = {{NuSTAR observation of {Ark} 564 reveals the variation of coronal temperature with flux}},
  journal      = {Monthly Notices of the Royal Astronomical Society},
  year         = {2020},
  volume       = {492(2)},
  pages        = {3041–3046},
  doi          = {10.1093/mnras/staa067},
}

@article{GuCao2009_LLAGN,
  author       = {Gu, Minfeng and Cao, Xinwu},
  title        = {{The anticorrelation between the hard {X‑ray} photon index and the {Eddington} ratio in low‑luminosity active galactic nuclei}},
  journal      = {Monthly Notices of the Royal Astronomical Society},
  year         = {2009},
  volume       = {399},
  pages        = {349‑356},
  doi          = {10.1111/j.1365-2966.2009.15277.x},
}

@article{Younes2011_LINERs,
  author       = {Younes, G. and Porquet, D. and Sabra, B. and Reeves, J. N.},
  title        = {{Study of {LINER} sources with broad H$\alpha$ emission. {X‑ray} properties and comparison to luminous {AGN} and {X‑ray} binaries}},
  journal      = {Astronomy \& Astrophysics},
  volume       = {530},
  pages        = {A149},
  year         = {2011},
  doi          = {10.1051/0004-6361/201116806},
}

@article{JangGliozzi2014_LowAccretingAGN,
  author       = {Jang, I. and Gliozzi, M. and Hughes, C. and Titarchuk, L.},
  title        = {{Constraining black hole masses in low‑accreting {AGN} using {X‑ray} spectra}},
  journal      = {Monthly Notices of the Royal Astronomical Society},
  year         = {2014},
  volume       = {443},
  pages        = {72-85},
  doi          = {10.1093/mnras/stu1024},
}

@article{Trakhtenbrot2017_BASS_Gamma_Ledd,
  author       = {Trakhtenbrot, Benny and others},
  title        = {{BAT AGN Spectroscopic Survey (BASS) — VI. The $\Gamma_{x}$ - L/$L_{Edd}$ relation}},
  journal      = {Monthly Notices of the Royal Astronomical Society},
  year         = {2017},
  volume       = {470},
  pages        = {800‑814},
  doi          = {10.1093/mnras/stx1117},
}

@article{SchulzeWisotzki2010,
  author       = {Schulze, Andreas and Wisotzki, Lutz},
  title        = {{Low redshift AGN in the Hamburg/ESO Survey. II. The active black hole mass function and the distribution function of Eddington ratios}},
  journal      = {Astronomy \& Astrophysics},
  year         = {2010},
  volume       = {516},
  pages        = {A87},
  doi          = {10.1051/0004-6361/201014193},
}

@article{NetzerTrakhtenbrot2007,
  author       = {Netzer, Hagai and Trakhtenbrot, Benny},
  title        = {{Cosmological evolution of mass accretion rate and metalicity in active galactic nuclei}},
  journal      = {The Astrophysical Journal},
  year         = {2007},
  volume       = {654},
  number       = {2},
  pages        = {754--763},
  doi          = {10.1086/509650}
}

@article{panessa,
       author = {{Panessa}, Francesca and {Baldi}, Ranieri Diego and {Laor}, Ari and {Padovani}, Paolo and {Behar}, Ehud and {McHardy}, Ian},
        title = "{The origin of radio emission from radio-quiet active galactic nuclei}",
      journal = {Nature Astronomy},
         year = 2019,
       volume = {3},
        pages = {387-396},
          doi = {10.1038/s41550-019-0765-4},
}

@article{kawamuro,
       author = {{Kawamuro}, Taiki and {Ueda}, Yoshihiro and {Tazaki}, Fumie and {Ricci}, Claudio and {Terashima}, Yuichi},
        title = "{Suzaku Observations of Moderately Obscured (Compton-thin) Active Galactic Nuclei Selected by Swift/BAT Hard X-ray Survey}",
      journal = {ApJS},
         year = 2016,
       volume = {225},
       number = {1},
        pages = {14},
          doi = {10.3847/0067-0049/225/1/14},
}

@ARTICLE{haardt,
       author = {{Haardt}, F. and {Maraschi}, L.},
        title = "{A Two-Phase Model for the X-Ray Emission from Seyfert Galaxies}",
      journal = {ApJ},
         year = 1991,
       volume = {380},
        pages = {L51},
          doi = {10.1086/186171},
}

@INPROCEEDINGS{2017xru..conf...84G,
       author = {{Gabriel}, C.},
        title = "{XMM-Newton Science Analysis System (SAS): medium and long term strategy}",
    booktitle = {The X-ray Universe 2017},
         year = 2017,
       editor = {{Ness}, Jan-Uwe and {Migliari}, Simone},
        month = oct,
        pages = {84},
       adsurl = {https://ui.adsabs.harvard.edu/abs/2017xru..conf...84G}
}

@ARTICLE{Duras2020,
       author = {{Duras}, F. and {Lusso}, E. and {Risaliti}, G. and {Bisogni}, S. and {Civano}, F. and {Nardini}, E. and {Vignali}, C. and {Zappacosta}, L. and {Bischetti}, M. and {Brusa}, M. and {Martocchia}, S. and {Piconcelli}, E. and {Salvestrini}, F. and {Severgnini}, P. and {Tombesi}, F. and {Vietri}, G.},
        title = "{The WISSH quasars project. IX. The 2-10 keV bolometric correction for hyper-luminous quasars}",
      journal = {aap},
         year = 2020,
        month = apr,
       volume = {636},
          eid = {A73},
        pages = {A73},
          doi = {10.1051/0004-6361/201937141},
archivePrefix = {arXiv},
       eprint = {2002.09457},
 primaryClass = {astro-ph.GA},
       adsurl = {https://ui.adsabs.harvard.edu/abs/2020A&A...636A..73D}
}

@article{shehata2021redshift,
  title={Redshift evolution of X-ray spectral index of quasars observed by XMM-NEWTON/SDSS},
  author={Shehata, SH M and Misra, R and Osman, AMI and Shalabiea, OM and Hayman, ZM},
  journal={Journal of High Energy Astrophysics},
  volume={31},
  pages={37--43},
  year={2021},
  publisher={Elsevier}
}

@article{ Laurenti,
	author = {{Laurenti, M.} and {Tombesi, F.} and {Vagnetti, F.} and {Piconcelli, E.} and {Guainazzi, M.} and {Vignali, C.} and {Paolillo, M.} and {Middei, R.} and {Bongiorno, A.} and {Zappacosta, L.}},
	title = {Investigating the nuclear properties of highly accreting active galactic nuclei with XMM-Newton},
	DOI= "10.1051/0004-6361/202449147",
	journal = {Astronomy \& Astrophysics},
	year = 2024,
	volume = 689,
	pages = "A337",
}

@article{Jana,
  title={{Coronal properties of low-accreting AGNs using Swift, XMM--Newton, and NuSTAR observations}},
  author={Jana, Arghajit and Chatterjee, Arka and Chang, Hsiang-Kuang and Nandi, Prantik and Rubinur, K and Kumari, Neeraj and Naik, Sachindra and Safi-Harb, Samar and Ricci, Claudio},
  journal={Monthly Notices of the Royal Astronomical Society},
  volume={524},
  number={3},
  pages={4670--4687},
  year={2023},
}

@article{Narayan,
       author = {{Narayan}, Ramesh and {Yi}, Insu},
        title = "{Advection-dominated Accretion: Underfed Black Holes and Neutron Stars}",
      journal = {ApJ},
         year = 1995,
       volume = {452},
        pages = {710},
          doi = {10.1086/176343},
}

@ARTICLE{Kelly,
       author = {{Kelly}, Brandon C. and {Vestergaard}, Marianne and {Fan}, Xiaohui and {Hopkins}, Philip and {Hernquist}, Lars and {Siemiginowska}, Aneta},
        title = "{Constraints on Black Hole Growth, Quasar Lifetimes, and Eddington Ratio Distributions from the SDSS Broad-line Quasar Black Hole Mass Function}",
      journal = {ApJ},
         year = 2010,
       volume = {719},
       number = {2},
        pages = {1315-1334},
          doi = {10.1088/0004-637X/719/2/1315},
}

@ARTICLE{Lusso,
       author = {{Lusso}, E. and {Comastri}, A. and {Simmons}, B.~D. and {Mignoli}, M. and {Zamorani}, G. and {Vignali}, C. and {Brusa}, M. and {Shankar}, F. and {Lutz}, D. and {Trump}, J.~R. and {Maiolino}, R. and {Gilli}, R. and {Bolzonella}, M. and {Puccetti}, S. and {Salvato}, M. and {Impey}, C.~D. and {Civano}, F. and {Elvis}, M. and {Mainieri}, V. and {Silverman}, J.~D. and {Koekemoer}, A.~M. and {Bongiorno}, A. and {Merloni}, A. and {Berta}, S. and {Le Floc'h}, E. and {Magnelli}, B. and {Pozzi}, F. and {Riguccini}, L.},
        title = "{Bolometric luminosities and Eddington ratios of X-ray selected active galactic nuclei in the XMM-COSMOS survey}",
      journal = {Monthly Notices of the Royal Astronomical Society},
         year = 2012,
       volume = {425},
       number = {1},
        pages = {623-640},
          doi = {10.1111/j.1365-2966.2012.21513.x},
}

@ARTICLE{Steinhardt,
       author = {{Steinhardt}, Charles L. and {Elvis}, Martin},
        title = "{The quasar mass-luminosity plane - I. A sub-Eddington limit for quasars}",
      journal = {Monthly Notices of the Royal Astronomical Society},
         year = 2010,
       volume = {402},
       number = {4},
        pages = {2637-2648},
          doi = {10.1111/j.1365-2966.2009.16084.x},
}

@ARTICLE{Shen,
       author = {{Shen}, Yue},
        title = "{The mass of quasars}",
      journal = {Bulletin of the Astronomical Society of India},
         year = 2013,
        month = mar,
       volume = {41},
       number = {1},
        pages = {61-115},
          doi = {10.48550/arXiv.1302.2643},
}

@article{Lusso_2016,
author = {Lusso, E. and Risaliti, G.},
title = {THE TIGHT RELATION BETWEEN {X-ray} AND ULTRAVIOLET LUMINOSITY OF QUASARS},
journal = {The Astrophysical Journal},
year = {2016},
publisher = {The American Astronomical Society},
volume = {819},
number = {2},
pages = {154},
doi = {10.3847/0004-637X/819/2/154},
}

@ARTICLE{Laha,
AUTHOR={Laha, Sibasish  and Ricci, Claudio  and Mather, John C.  and Behar, Ehud  and Gallo, Luigi  and Marin, Frederic  and Mbarek, Rostom  and Hankla, Amelia },
TITLE={X-ray properties of coronal emission in radio quiet active galactic nuclei},
JOURNAL={Frontiers in Astronomy and Space Sciences},
VOLUME={Volume 11 - 2024},
YEAR={2025},
URL={https://www.frontiersin.org/journals/astronomy-and-space-sciences/articles/10.3389/fspas.2024.1530392},
DOI={10.3389/fspas.2024.1530392},
ISSN={2296-987X},
}

@article{Done,
    author = {Done, Chris and Gierliński, Marek and Kubota, Aya},
    title = {Modelling the behaviour of accretion flows in {X-ray} binaries},
    journal = {The Astronomy and Astrophysics Review},
    year = {2007},
    volume = {15},
    pages = {1-66},
    doi ={10.1007/s00159-007-0006-1},
}

@article{Laurenti2022,
	author = {{Laurenti, M.} and {Piconcelli, E.} and {Zappacosta, L.} and {Tombesi, F.} and {Vignali, C.} and {Bianchi, S.} and {Marziani, P.} and {Vagnetti, F.} and {Bongiorno, A.} and {Bischetti, M.} and {del Olmo, A.} and {Lanzuisi, G.} and {Luminari, A.} and {Middei, R.} and {Perri, M.} and {Ricci, C.} and {Vietri, G.}},
	title = {X-ray spectroscopic survey of highly accreting {AGN}},
	DOI= "10.1051/0004-6361/202141829",
	journal = {Astronomy \& Astrophysics},
	year = 2022,
	volume = 657,
	pages = "A57",
}

@article{wu2022catalog,
  title={A catalog of quasar properties from sloan digital Sky survey data release 16},
  author={Wu, Qiaoya and Shen, Yue},
  journal={The Astrophysical Journal Supplement Series},
  volume={263},
  number={2},
  pages={42},
  year={2022},
  publisher={IOP Publishing}
}

@article{jiang2007radio,
  title={The radio-loud fraction of quasars is a strong function of redshift and optical luminosity},
  author={Jiang, Linhua and Fan, Xiaohui and Ivezi{\'c}, {\v{Z}}eljko and Richards, Gordon T and Schneider, Donald P and Strauss, Michael A and Kelly, Brandon C},
  journal={The Astrophysical Journal},
  volume={656},
  number={2},
  pages={680},
  year={2007},
  publisher={IOP Publishing}
}


\end{document}